\documentclass[final,5p,times,twocolumn]{elsarticle}

\usepackage{amsmath,amsfonts,upgreek}
\usepackage{algorithm}
\usepackage{algpseudocode}
\usepackage{hyperref}
\journal{Nuclear Inst. and Methods in Physics Research, A}

\begin{document}

\begin{frontmatter}


\title{HitPix3 – a new HV-CMOS active pixel sensor for ion beam therapy}

\author[kit_ipe]{Hui Zhang\fnref{label1}}
\ead{hui.zhang@kit.edu}

\affiliation[kit_ipe]{organization={Institute for Data Processing and Electronics (IPE), Karlsruhe Institute of Technology},
            city={Karlsruhe},
            postcode={76131}, 
            country={Germany}}

\author[kit_etp]{Bogdan Topko\corref{cor1}\fnref{label1}}
\ead{bogdan.topko@kit.edu}
\cortext[cor1]{Corresponding author}
\fntext[label1]{Co-first authors \\ \textit{Preprint submitted to Nuclear Inst. and Methods in Physics Research, A September 26, 2024; Revised January 7, 2025; Revised June 14, 2025; Revised September 10, 2025}}

\author[kit_ipe]{Ivan Peri\'c}

\affiliation[kit_etp]{organization={Institute of Experimental Particle Physics (ETP), Karlsruhe Institute of Technology},
            city={Karlsruhe},
            postcode={76131}, 
            country={Germany}}

\begin{abstract}
Particle therapy is a well-established clinical treatment of tumors and so far, more than one hundred centers are in operation around the world. High accuracy on position and dose rate in beam monitoring are major cornerstones of clinical success in particle therapy.
A high voltage CMOS (HV-CMOS) monolithic active pixel sensor was developed for this beam monitoring system. The HV-CMOS technology has demonstrated many advantages over other technologies in dealing with radiation tolerance. This HV-CMOS sensor was produced with a 180~nm commercial technology on high-resistivity substrate. The HitPix sensor features several specific design details for operation in a high-intensity ion beam environment: hit-counting pixels, on-sensor projection calculation, radiation tolerant design and frame-based readout. It has a 9775~$\upmu$m~$\times$~10110~$\upmu$m sensor area with 200~$\upmu$m$~\times~$200~$\upmu$m pixel size.
The new HitPix3 has a number of improvements, including a modified in-pixel amplifier, on-sensor calculation of beam profile projection in two dimensions, and in-pixel threshold tuning capacity. The functionality of these features was confirmed in laboratory tests of unirradiated HitPix3 sensors, making the HitPix3 an important step in developing a sensor for use in ion beam monitoring.

\end{abstract}

\begin{keyword}
HV-CMOS sensor \sep monolithic integrated circuits \sep radiation-hard sensor \sep high-rate beam monitoring \sep pixel hit counting \sep instrumentation for hadron therapy
\end{keyword}

\end{frontmatter}

\section{Introduction}
The HitPix3 is a new high voltage CMOS (HV-CMOS) sensor based on hit counting electronics and frame-based readout which aims to be used as a beam monitor for ion beam therapy. The sensor is customized to the demand of the beam monitoring system at the Heidelberg Ion Beam Therapy center (HIT) and is part of the R\&D program targeting a 2~cm~$\times$~2~cm sensor foreseen to be assembled on a detection plane with a 26~cm~$\times$~26~cm sensitive area~\cite{Dierlamm23}.

The beam monitoring system must satisfy certain requirements to ensure a patient's safe treatment and defined by HIT. These requirements are detailed below.

Firstly, the system must provide the beam position and size (FWHM) to a fast control interface in real time with a 100~$\upmu$s latency~\cite{Dierlamm23}. This ensures that the irradiation will immediately stop if the beam parameters deviate from the treatment plan. The measured particle hit rate together with a known beam energy could be used to track the applied dose. Secondly, the accuracy of the measured beam position and size must be better than 200~$\upmu$m and 400~$\upmu$m, respectively~\cite{Dierlamm23}. Furthermore, the system should work for beam intensities up to 10$^{10}$~s$^{-1}$ for protons, be tolerant to 1~MeV neutron equivalent fluences up to 10$^{15}$~cm$^{-2}$ per year and to magnetic fields (for MRI-guided ion beam therapy)~\cite{Dierlamm23}. All these requirements should be satisfied while operating at a temperature of 0~$^{\circ}$C~to ensure high detection efficiency after significant radiation damage~\cite{Topko24}. The system should cover a 200~mm~$\times$~200~mm treatment field~\cite{HABERER2004S186} with an adjustable beam size (FWHM) in the range from 4~mm to 30~mm at the isocenter of HIT beamline~\cite{SCHOEMERS201592}.

An overview of the beam monitoring system requirements, limitations of the current system (based on multi-wire proportional chambers) for MRI-guided ion beam therapy, and alternative beam monitoring concepts and devices can be found in~\cite{Dierlamm23}.

The monolithic active pixel sensor was implemented with a commercial 180~nm HV-CMOS technology (TSI Semiconductors) on high resistivity substrate. The HV-CMOS process allows producing sensor and readout electronics in the same die. It achieves enhanced radiation tolerance by using high external reverse bias voltage (up to 150~V) to enlarge the depletion region and accelerate the charge collection speed.

The previous versions of these HV-CMOS sensors -- HitPix1 and HitPix2 were successfully developed and tested~\cite{Dierlamm23, Topko24, Weber2022}.
They have proven to be a promising candidate for a beam monitoring system. 

The sensor is adapted for the beam position and size resolution requirements yielding an overall area with  9775~$\upmu$m~$\times$~10110~$\upmu$m, active area of 9600~$\upmu$m~$\times$~9600~$\upmu$m (93.3\%~of~the~total~area), and individual pixel size of~200~$\upmu$m $\times$ 200 $\upmu$m. This large active area is achieved by constraining the peripheral electronics to an area of 9600~$\upmu$m~$\times$~420~$\upmu$m (4.1\%~of~the~total~area) at the bottom of the sensor. The area between the guard ring and sensor edges occupies the rest 2.6\%~of~the~total~area. Readout concepts used in the sensor design allow to form multi-sensor assemblies and fulfill the large sensitive area requirement. The hit counting capability and the built-in $x$ projection onto columns aim for processing high particle rates and fast readout at minimum latency (100~$\upmu$s).

Several new designs were implemented and fabricated in HitPix3. These include a modified amplifier, a second hit counting projection (in $y$), which acts onto rows, in-pixel discriminator threshold tuning, and a faster readout concept.

\section{A beam monitor based on an HV-CMOS Sensor}
Particle sensors made of silicon based material are widely used in experimental particle physics and were proposed for usage in radiotherapy for many years. However, most silicon tracking detector implementations have their limitations with respect to high-rate ion beam monitoring. The hybrid pixel detectors have additional insensitive material of the readout ASIC, which can affect the beam position and size, due to increased particle scattering~\cite{Rossi2006, garcia-sciveres_review_2018}. The Monolithic Active Pixel Sensors (MAPS), based on low-voltage CMOS technology, and double-sided silicon strip detectors have limited capabilities for counting particles at high rates, due to the slow charge collection via diffusion in MAPS, or the "ghost track" ambiguities in strip detectors~\cite{garcia-sciveres_review_2018, MAGER2016434, SEIDEL20191}. The typical slow charge collection time of these MAPS sensors require an integration time in the order of milliseconds~\cite{FLYNN2022166703} and do not meet the timing requirements of the HIT beam monitoring system, which should provide an estimation of the beam parameters in a few hundred microseconds. Furthermore, the typical radiation tolerance of the low-voltage MAPS is limited to 1-MeV neutron equivalent fluences on the order of 10$^{13}$~cm$^{-2}$~\cite{garcia-sciveres_review_2018, MAGER2016434}. The Low Gain Avalanche Detectors (LGADs) provide excellent time resolution suitable for counting particles at high rates, as reported in~\cite{Vignati_2022}. However, their gain parameter can degrade at a proton fluence of 10$^{12}$~cm$^{-2}$ as reported in~\cite{Sacchi_2020}. Furthermore, the LGAD requires a separate readout ASIC, leading to the same challenges mentioned with hybrid detectors. These can limit the application of LGADs for direct beam monitoring.

\begin{figure}[!hbt]
\centering
\includegraphics[width=2.6in]
{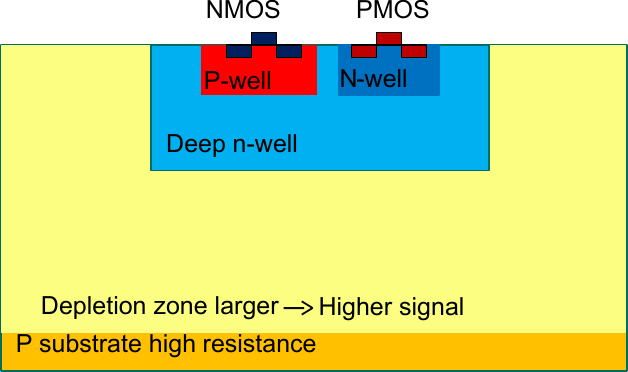}
\caption{Cross-section illustrating the HV-CMOS technology~\cite{Zhang2022_phd}.}
\label{fig_cs_hvcmos}
\end{figure}

In contrast, the high voltage CMOS pixel sensors~\cite{Peric2021, PERIC2007876} have several advantages: they employ commercial HV-CMOS sensor production technologies and allow combining the detector with readout electronics in one thin (down to 50~$\upmu$m) sensor, reducing the material budget. The cross-section of the HV-CMOS technology is shown in~Fig.~\ref{fig_cs_hvcmos}. The deep n-well and the p-type substrate form a diode that works as a detector. The electronics are embedded in a deep n-well. The external negative voltage applied between the bias contact and the p-type substrate generates a depletion zone and accelerates the charge collection by drift (achieved within nearly 1~ns). The possibility of charge trapping after irradiation is decreased. HV-CMOS sensor prototypes have shown a radiation tolerance up to 100~MRad of total ionizing dose and 1~$\times$~10$^{15}$~cm$^{-2}$ of 1 MeV neutron equivalent fluence with a detection efficiency of 99.7\% in beam tests~\cite{Weber2022, Benoit_2018}. Therefore, combining improved radiation tolerance with integrated electronics and small pixel sizes makes the HV-CMOS technology a promising candidate for an ion beam monitoring system.

\section{Sensor design}
\begin{figure}[!hbt]
\centering
\includegraphics[width=2.6in]
{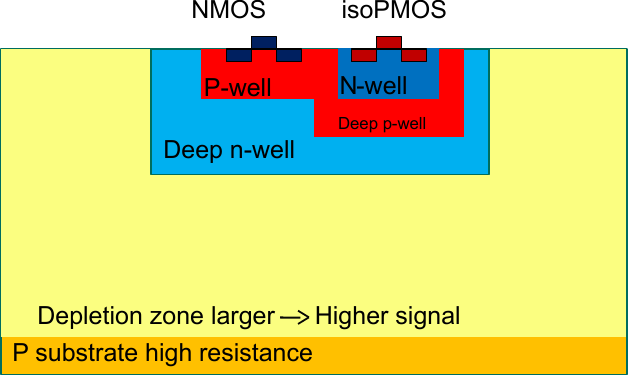}
\caption{Modified HV-CMOS sensor designed with a deep p-well implant~\cite{Zhang2022_phd}.}
\label{fig_modified_hvcmos}
\end{figure}

The HitPix3 was submitted with a 180~nm HV-CMOS process on a 370~Ohm~$\times$~cm resistivity substrate. 
The high-voltage process leads to a larger depletion region by higher reverse bias and higher resistivity substrate compared to the standard CMOS process.
This configuration reduces the sensor capacitance, ensures lower noise, and enhances the input signal for detecting charged particles.
The modified HV-CMOS sensor design is illustrated in~Fig.~\ref{fig_modified_hvcmos}.

\begin{figure}[!hbt]
\centering
\includegraphics[width=2.6in]
{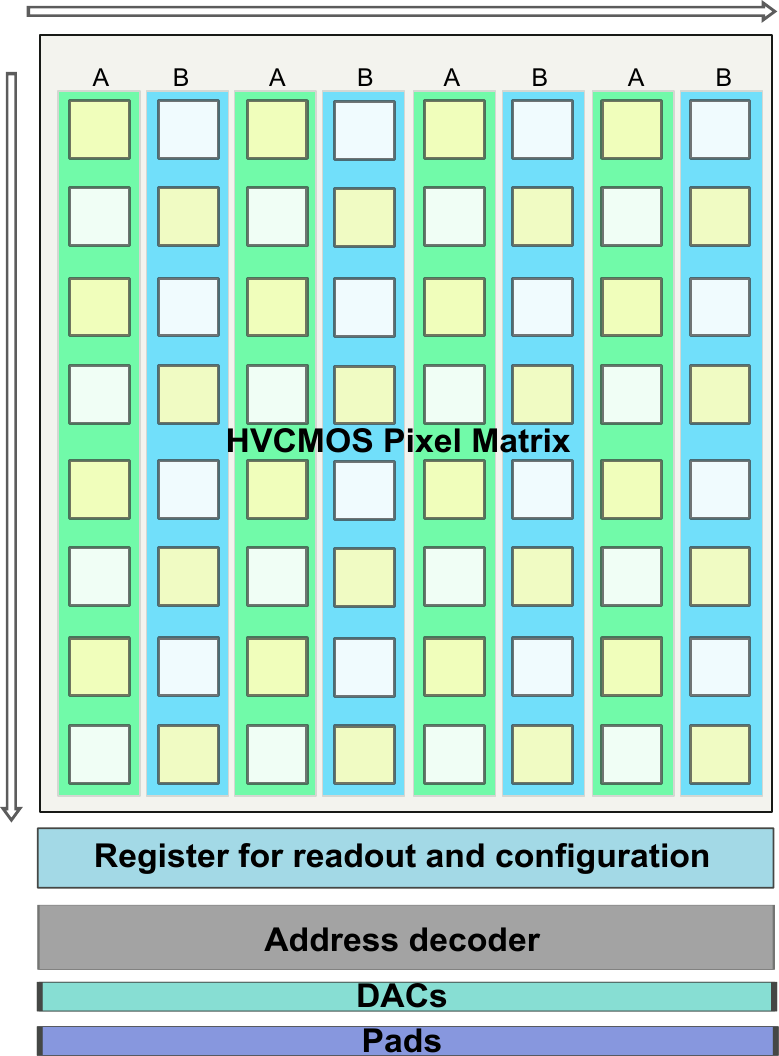}
\caption{Simplified block diagram of the HitPix3 sensor.}
\label{fig_chip_design}
\end{figure}

The n-wells hosting PMOS transistors are embedded in deep p-wells which isolate them from the deep n-wells ensuring charge collection. This isolation is important to avoid shorting of the sensor n-well with the supply voltage (V$_{dd}$) and to prevent capacitive cross-talk of digital signals to the sensor~\cite{Zhang2023}. The pixel electronics are implanted in the deep n-well. To maximize the radiation tolerance, radiation tolerant PMOS transistors, enclosed NMOS transistors and guard ring structures were implemented in the layout design. 
A simplified sensor diagram is sketched in~Fig.~\ref{fig_chip_design}. The sensor size is about 1~cm~$\times$~1~cm. It consists of 48~$\times$~48 pixels with 200~$\upmu$m$~\times~$200~$\upmu$m pixel size, a register for readout and configuration, address decoder, bias voltage-DAC (VDAC) and pads.

The pixel size allows to integrate the entire detector signal processing chain -- including an amplifier and hit-counting electronics -- directly within each pixel. There are two readout modes: counter mode and adder mode. The former is used for beam diagnosis and the latter is used for quasi-instantaneous readout.
Additionally, two different column types (A and B) in the pixel matrix are introduced. In the columns of type B, the counter values of the pixels are added vertically. The counters of the pixels composing A-type columns are added row-wise. This is new with respect to the previous prototypes (HitPix1 and HitPix2), which only provide a one-dimensional projection, i.e. in the $x$-direction (column-wise).

\section{Pixel electronics} \label{pixel_elec_section}
\begin{figure}[!hbt]
\centering
\includegraphics[width=3.4in]
{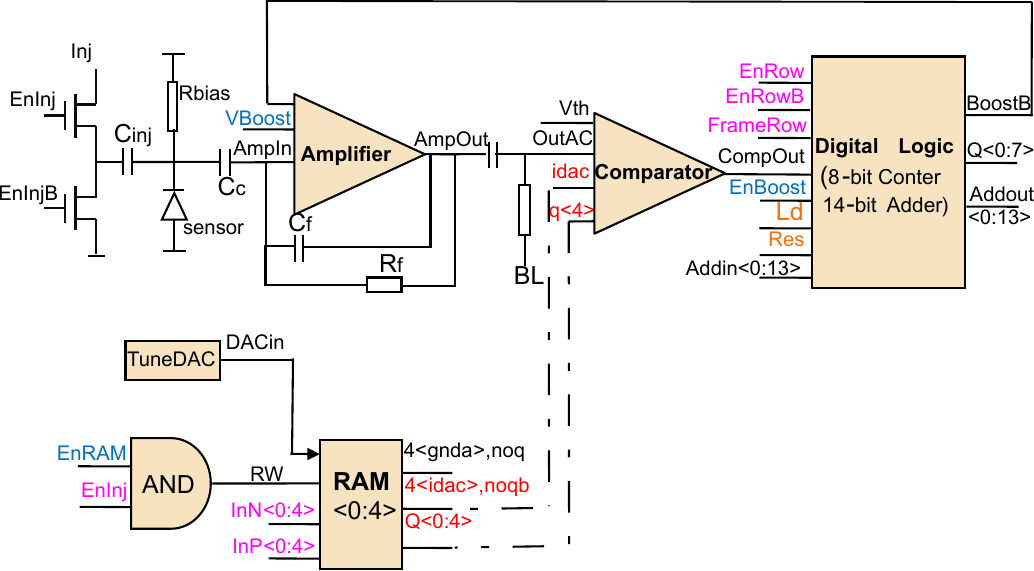}
\caption{Schematic of pixel electronics.}
\label{fig_pixel_electronics}
\end{figure}

The pixel electronics consist of a charge sensitive amplifier (CSA), comparator, 5-bit RAM with tune DAC and digital logic, shown in~Fig.~\ref{fig_pixel_electronics}.

The signal detection is based on the following principle: When a particle hits a pixel, the signal charge (ionization electrons) is collected by the positively biased n-well, thus a small voltage drop is generated and detected by the amplifier. The voltage step pulse at the output of the amplifier is proportional to the amount of the charge collected in the sensor and inversely proportional to the feedback capacitance C$_f$($\sim$1.6~fF), which is discharged by a continuous reset R$_f$.

\subsection{Charge sensitive amplifier}
A folded cascode topology and a source follower were used in the CSA~\cite{Peric2021}. The input PMOS transistor of the CSA is AC coupled to the n-well via a coupling capacitance C$_c$. Generally, PMOS transistors have lower flicker noise and higher radiation hardness than NMOS transistors~\cite{FACCIO20081000}.  Each pixel has a corresponding capacitive injection circuit to generate test signals and allow fast testing and debugging. This circuit utilizes the discharge of a capacitor C$_{inj}$ to inject a defined test charge into a specified pixel readout circuit. The CSA output of pixels in the row closest to the sensor peripheral electronics is connected to a test pad, which allow for direct measurement of the CSA output signals. 
The current through the CSA is 5~$\upmu$A and through the comparator is 2~$\upmu$A.
The total power consumption per pixel is below 10~$\upmu$W.

\subsection{TuneDAC}
The comparator is based on a differential amplifier to compare an input signal from the CSA to a threshold voltage V$_{th}$. The 8-bit threshold setting (VDAC.Vth) generated by the VDAC is global and applied to all pixels.

Previous versions of HitPix had only one global VDAC.Vth value, but this had limitations. When considering the non-uniformity of the noise across the whole matrix and the pixel-to-pixel thresholds dispersion, the minimum threshold value was constrained by the noise level.
Thus, the wider the threshold distribution, the larger the voltage must be set to ensure that all pixels have local thresholds above the noise level. Simultaneously, care has to be taken to avoid deteriorating the detection efficiency consecutive to a too high threshold as compared to the smallest signal amplitudes. This is a problem after irradiation since the signal amplitude is decreased and the noise is increased. 

Therefore, in the HitPix3 design, a local adjustment with a 4-bit tune DAC was implemented to cope with a wide threshold dispersion. The logic AND output of {\fontfamily{qcr}\selectfont EnRAM} and {\fontfamily{qcr}\selectfont EnInj} is used to write the pixel Q$\langle$0:4$\rangle$ RAM cell connected to the tune DAC. With the tune DAC, the threshold can be adjusted individually for each pixel. Thus, a small threshold could be ensured, which is critical for detection efficiency. Moreover, noisy pixels can be masked by bit Q$\langle$4$\rangle$, which becomes essential for operating a radiation damaged sensor in adder mode.

\subsection{Feedback Boost}
In previous HitPix iterations it was observed that, for a high rate of highly ionizing particles, the baseline voltage is affected and the hit detection performance deteriorates~\cite{Dierlamm23, Topko24}.

\begin{figure}[!hbt]
\centering
\includegraphics[width=3.0in]
{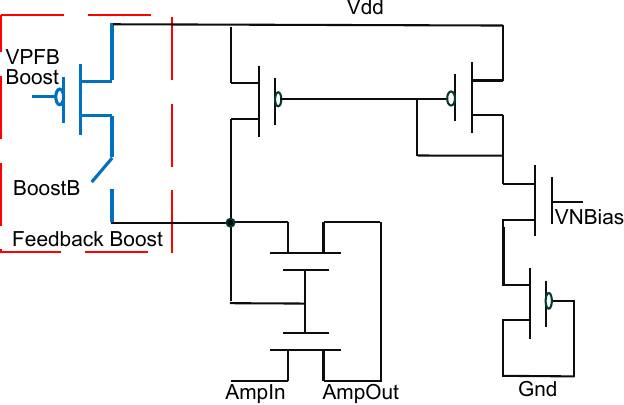}
\caption{Feedback circuit with the boost feature (red box).}
\label{fig_feedback_boost}
\end{figure}

To overcome this problem, the feedback circuits in the HitPix3 were modified with a "feedback boost" feature~\cite{C_Zhang2021_phd} (as shown in~Fig.~\ref{fig_feedback_boost}). The feedback boost feature enables the baseline to be restored faster via two signals ({\fontfamily{qcr}\selectfont VPFBBoost} and {\fontfamily{qcr}\selectfont BoostB}) that support the amplifier feedback circuit. This pathway is enabled with  {\fontfamily{qcr}\selectfont BoostB}, a logical NAND of the comparator output and the enable signal. When {\fontfamily{qcr}\selectfont BoostB} = 1, the switch is turned on and more current flows into the feedback circuits between {\fontfamily{qcr}\selectfont AmpIn} and {\fontfamily{qcr}\selectfont AmpOut}.

\subsection{Counter and adder}
\begin{figure}[!hbt]
\centering
\includegraphics[width=3.3in]
{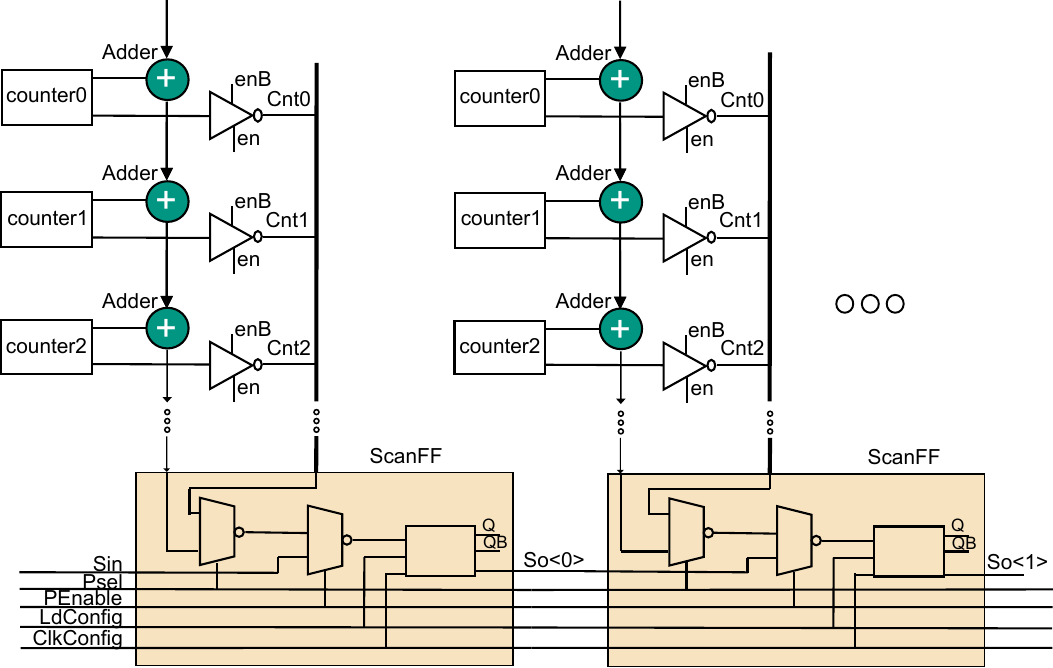}
\caption{Illustration of the hit counting and adding features of HitPix3 with a scan flip-flop and shift register.}
\label{fig_ill_cnt_add}
\end{figure}

The output of the comparator is fed into the digital logic with an 8-bit radiation-hard ripple counter and a 14-bit adder located inside each pixel. The 8-bit counter values and the 14-bit adder values are transferred into the readout register. The digital logic is enclosed in a guard ring. One frame can be read out either through counter mode (slow) or adder mode (fast). The hit counting and adding features of HitPix3 with a scan flip-flop and shift register are shown in~Fig.~\ref{fig_ill_cnt_add}.

The counter value range of each pixel goes from 0 to 255 and the maximal number of hits in one column is 12240 (48~$\times$~255). Each time a hit is detected, the counter value is increased by one. Then the value can be selected by writing its address into the Parallel In Serial Out (PISO) register. The address decoder defines which row to read. The counters in one column share the readout lines. In order to read all counter values from one column, all rows need to be selected and read out one by one.

An asynchronous 14-bit adder (2$^{14}$~=~16384) is located inside of the same guard ring as the counter. The adder reduces the data size and readout time by pre-processing data on the sensor. This is achieved by summing the counter value of the corresponding pixel cell with the value from the preceding adder. For a given direction the value of the last pixel holds the sum of all counters. The adder readout can be selected in the address decoder and loaded into the PISO while the counter works. If adders mode is used, the beam projection information can be read out in just one readout cycle (48 times faster than counter mode), which is much more suitable for continuous beam monitoring.

\begin{figure}[!hbt]
\centering
\includegraphics[width=2.8in]
{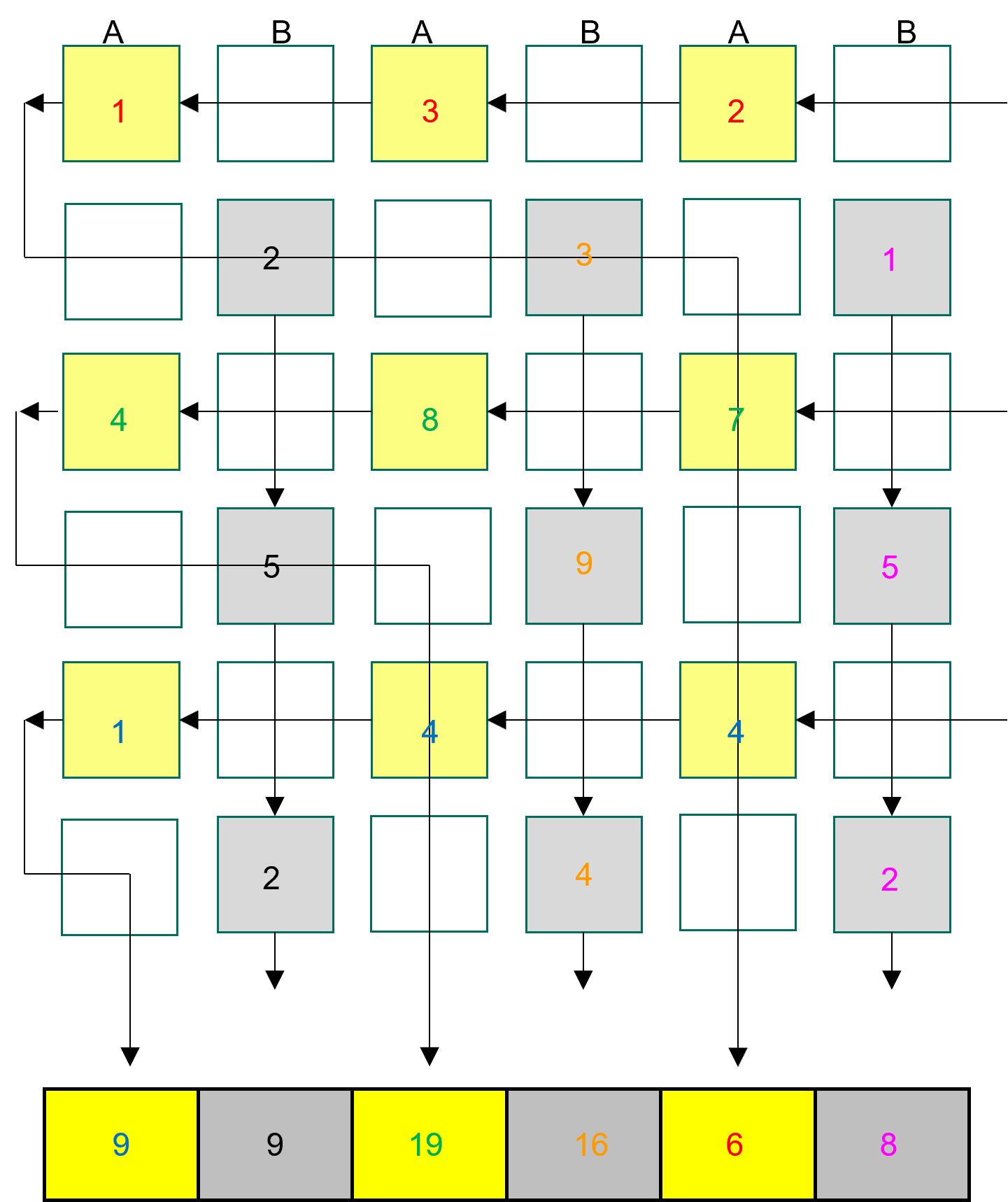}
\caption{The connection of counter values in adder mode for columns A and B. With the adders, two-dimensional (2D) projection information is available. With the sum information the user can decide which columns and rows should be read out in detail.}
\label{fig_proj_readout}
\end{figure}

As mentioned earlier, in the previous designs, the counter values can only be summed in the $x$-direction (column-wise).
In HitPix3, counter values can be summed up in both directions. 
Figure~\ref{fig_proj_readout} illustrates the working principle of the adder mode. 

The pixels are numbered from top to bottom and from left to right, starting with 0. 
In the case of columns B, the counter values of all odd pixels (shown as gray boxes in Fig.~\ref{fig_proj_readout}) are summed, and the result is transmitted to the readout register. 
In the case of columns A, the counters of all even pixels (shown as yellow boxes in Fig.~\ref{fig_proj_readout}) are summed row-wise and the result is also transmitted to the register.
In such a scheme, only one adder per pixel is required. The drawback is that about half of the pixels (shown as white boxes in Fig.~\ref{fig_proj_readout}) are not included when summing. However, this is not a problem because the beam profile is much lager than the size of an individual pixel.

\subsection{Pixel layout}
\begin{figure}[!hbt]
\centering
\includegraphics[width=2.8in]
{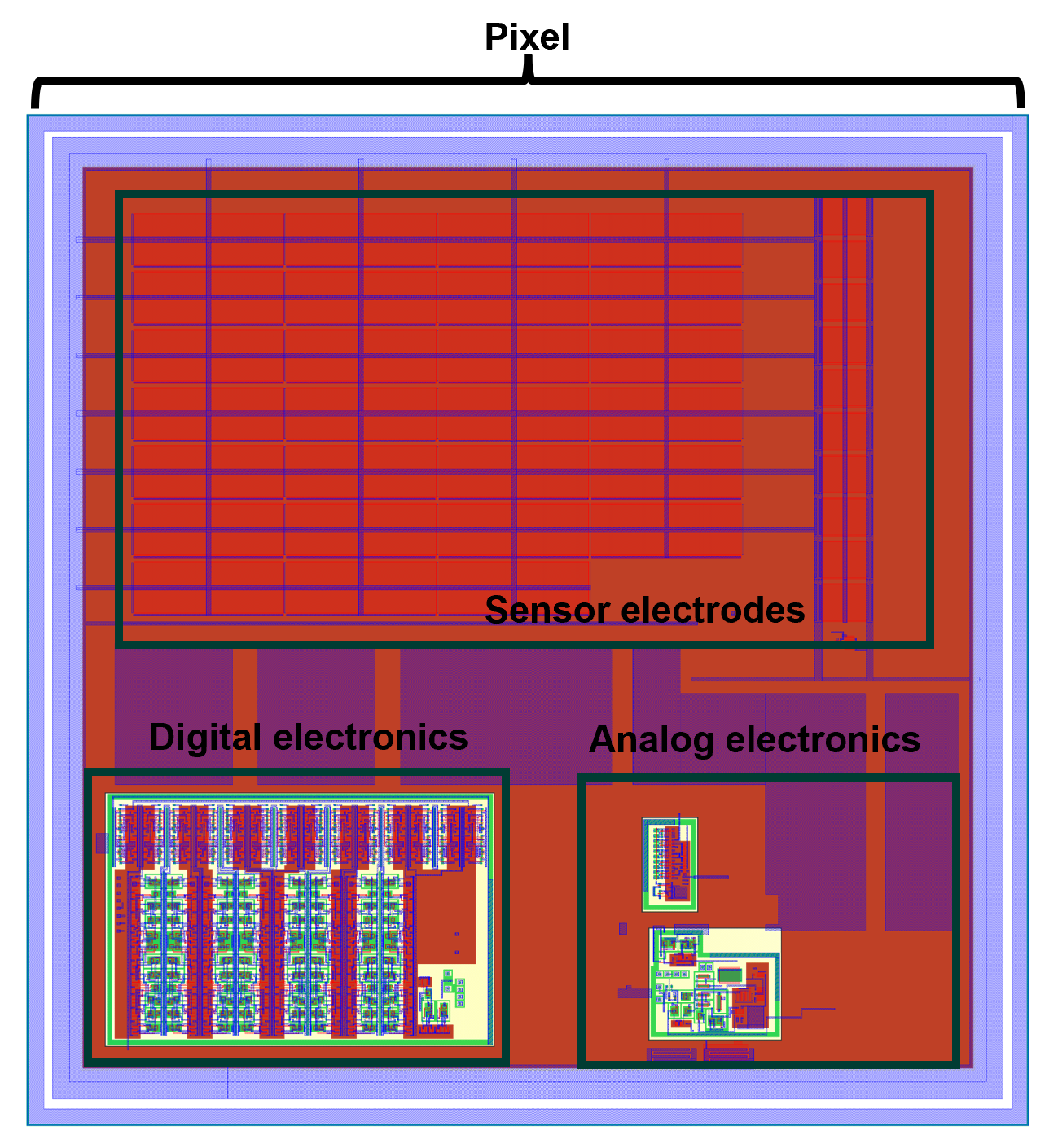}
\caption{The HitPix3 pixel layout (200~$\upmu$m$~\times~$200~$\upmu$m), comprising the sensor electrodes, the analog electronics (bottom right) and the digital electronics (bottom left).}
\label{fig_pixel_layout}
\end{figure}

The pixel layout is shown in~Fig.~\ref{fig_pixel_layout}. The pixel area is dominated (80\% of the area) by a deep n-well, which contains the sensor electrode.

The analog electronics, placed at the bottom right of the layout, include the CSA, the comparator, the bias and feedback circuits and the RAM with the tune DAC. 
The digital electronics (an 8-bit counter and 14-bit adder) can cause noise and were, therefore, placed at the bottom left of the layout, separated from the analog electronics.
The p- and n-wells with transistors are isolated from the deep n-well using deep p-wells (black lines in figure). 

\section{Sensor peripheral electronics}

The sensor periphery consists of registers, an address decoder, bias DACs and pads. All these elements occupy~4.1\% of the whole design surface, being concentrated in a side band of the sensor with~420~$\upmu$m~height and a width equal to the pixel array width, thereby minimizing the insensitive area.

\begin{figure}[!hbt]
\centering
\includegraphics[width=2.8in]
{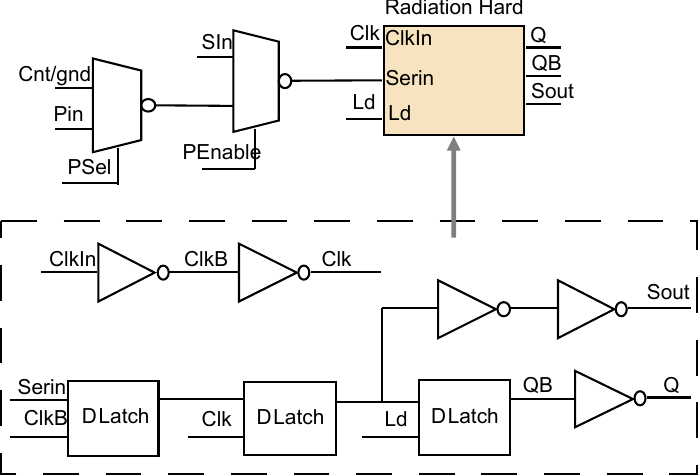}
\caption{Schematic diagram of the basic block of the register for readout and configuration, which stores a single bit.}
\label{fig_scan_ff_struct}
\end{figure}

Figure~\ref{fig_scan_ff_struct} displays the schematic diagram of the basic block of the readout and configuration register, which can store and transfer a single bit. With the first multiplexer and the signal {\fontfamily{qcr}\selectfont PSel} one can choose between the counter and adder value for the readout. With the second multiplexer and the signal {\fontfamily{qcr}\selectfont PEnable} one can choose between parallel load and shifting of bits. 
In the shifting mode, the register can also be used to load the configuration bits into the latch. 

The configuration bits are used for addressing pixel rows and enabling pixel wise readout and for writing into a pixel RAM.
To increase the readout speed, the signals for the register are provided through LVDS pads.

\section{Measurements}
\begin{figure}[!hbt]
\centering
\includegraphics[width=2.4in]{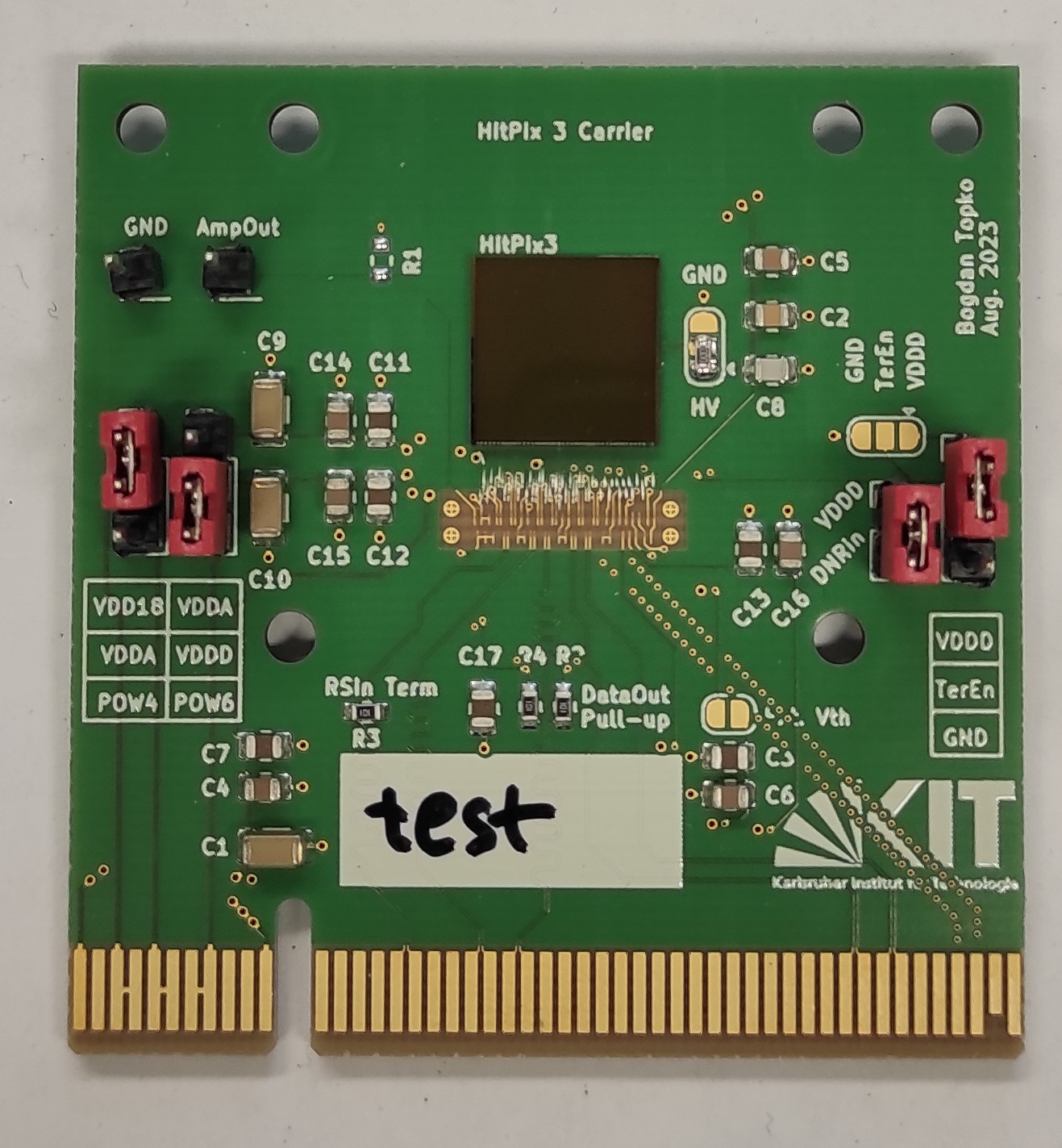}
\caption{The HitPix3 sensor assembled on the single sensor PCB carrier.}
\label{fig_phot_hitpix3}
\end{figure}

Several diced HitPix3 sensors were mounted and wire bonded on single-sensor PCB carriers to check all sensor functionalities in a laboratory environment. A photograph of the assembled sensor is shown in~Fig.~\ref{fig_phot_hitpix3}. The data acquisition system (DAQ) consists of an FPGA board for sensor control and readout, a GECCO FPGA extension board for LV, HV and control signal delivery, an oscilloscope, and a control PC. Detailed information about the HitPix DAQ system can be found in~\cite{Dierlamm23}. All measurements were performed at room temperature (+23~$^{\circ}$C). The readout speed was set to 180~Mbit/sec (close to the current readout firmware limit) to overcome the 100~$\upmu$s data transfer timing requirements in adder mode anticipated at HIT. The baseline voltage at the comparator input (BL in Fig.~\ref{fig_pixel_electronics}) was set to 1.0~V. The VDDA power supply, which affects the scale of the VDAC.Vth units, was set to 1.85~V, i.e. 1 DAC unit corresponds to $\frac{1}{256}\times1850$~mV~$=~7.2$~mV. The presented results were obtained using unirradiated sensors.

\subsection{I-V measurement}
\begin{figure}[!hbt]
\centering
\includegraphics[width=3.2in]{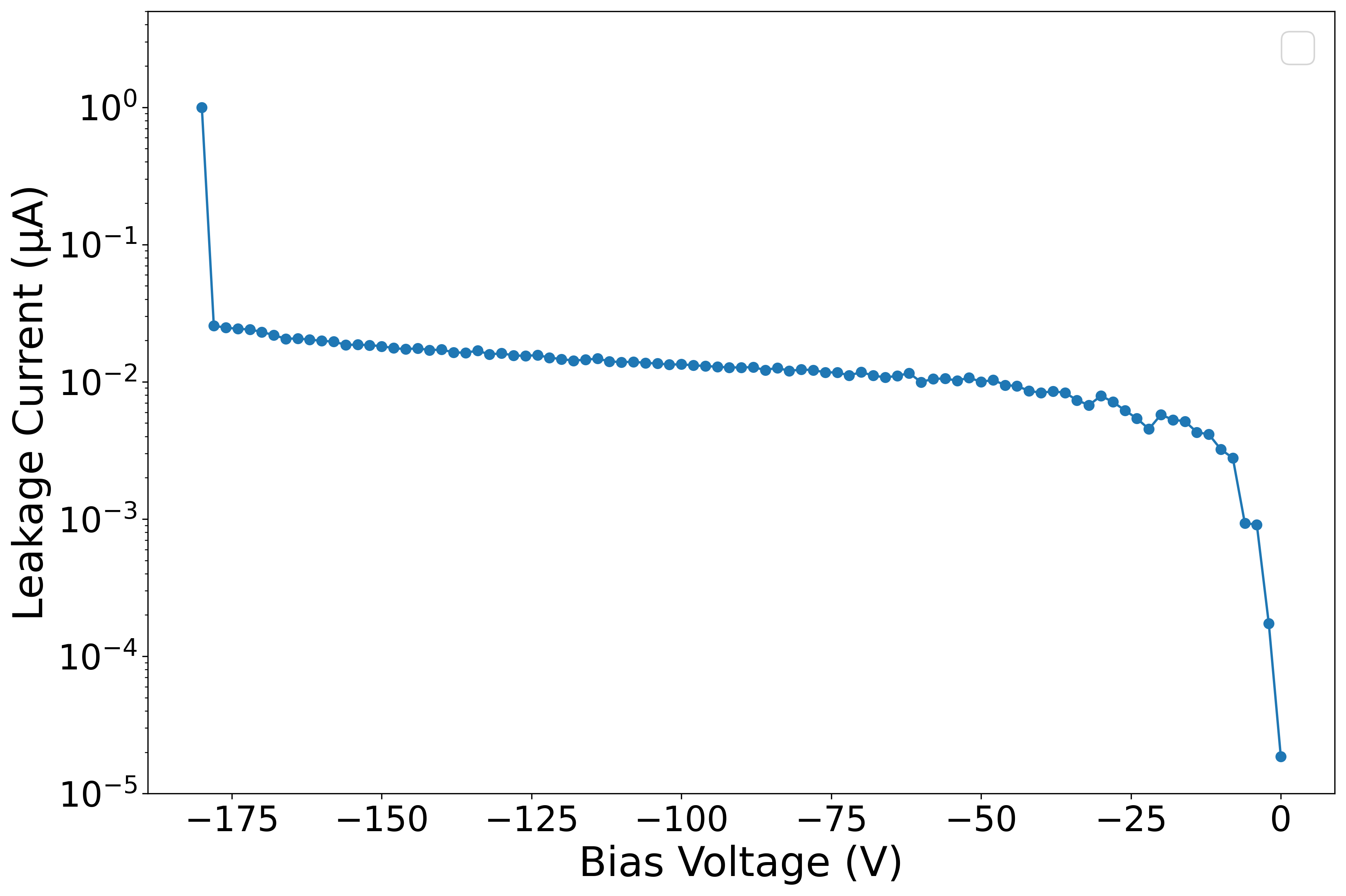}
\caption{Example of a measured I-V curve. The leakage current is plotted in absolute values.}
\label{fig_iv_curve}
\end{figure}

Current versus voltage (I-V) measurements were performed for the assembled sensors to check the average leakage current and breakdown voltage. The sensor was placed in a light-tight box and the bias voltage was decreased from 0~V to –180~V in 2~V steps. An example of a measured I-V curve is shown in~Fig.~\ref{fig_iv_curve}.

The I-V measurement confirmed that the designed sensor can hold bias voltages up to –180~V without breakdown and that the total leakage current of the sensor does not exceed 20~nA.

\subsection{Latency scan measurement}

The latency scan measurement tests the maximum readout speed of the sensor, which corresponds to the frequency of the shift register {\fontfamily{qcr}\selectfont Clk} signal. The measurement consists of the following steps: 1)~Generation of a random bit sequence with the size of the shift register, i.e. 14~$\times$~48~=~672~bits. 2)~Shifting the generated sequence into the sensor shift register. 3)~Shifting out the data from the shift register where each output bit is sampled by the FPGA with a controllable latency. 4)~Calculating the number of bits which were read out wrongly. 5)~Repeat the loop 10 times. By varying the latency value and the frequency value of the {\fontfamily{qcr}\selectfont Clk} signal the maximum readout speed can be determined. 
 
The previous HitPix2 sensor data shift register has a readout speed limited to 55~Mbit/s. 

The readout speed of the HitPix3 sensor was drastically increased by redesigning all signals according to the LVDS standard and using a one-phase clock for the shift register. It reaches the limitation of the current firmware, which is 190~Mbit/s.

\subsection{CSA calibration measurement}
As mentioned in section~\ref{pixel_elec_section}, the last row of the sensor matrix contains the testing output of the pixel CSA. Also, each pixel contains an injection circuit for testing the pixel electronics and optimizing the pixel comparator thresholds (more details may be found in section~\ref{pix_th_tuning}). A Sr-90 radioactive source was positioned above the last row of the sensor matrix. The output signals of the selected CSA were recorded with the oscilloscope and analyzed offline. The bias voltage was set to –90~V, which corresponds to a depletion depth of around 55~$\upmu$m.

\begin{figure}[!hbt]
\centering
\includegraphics[width=3.2in]{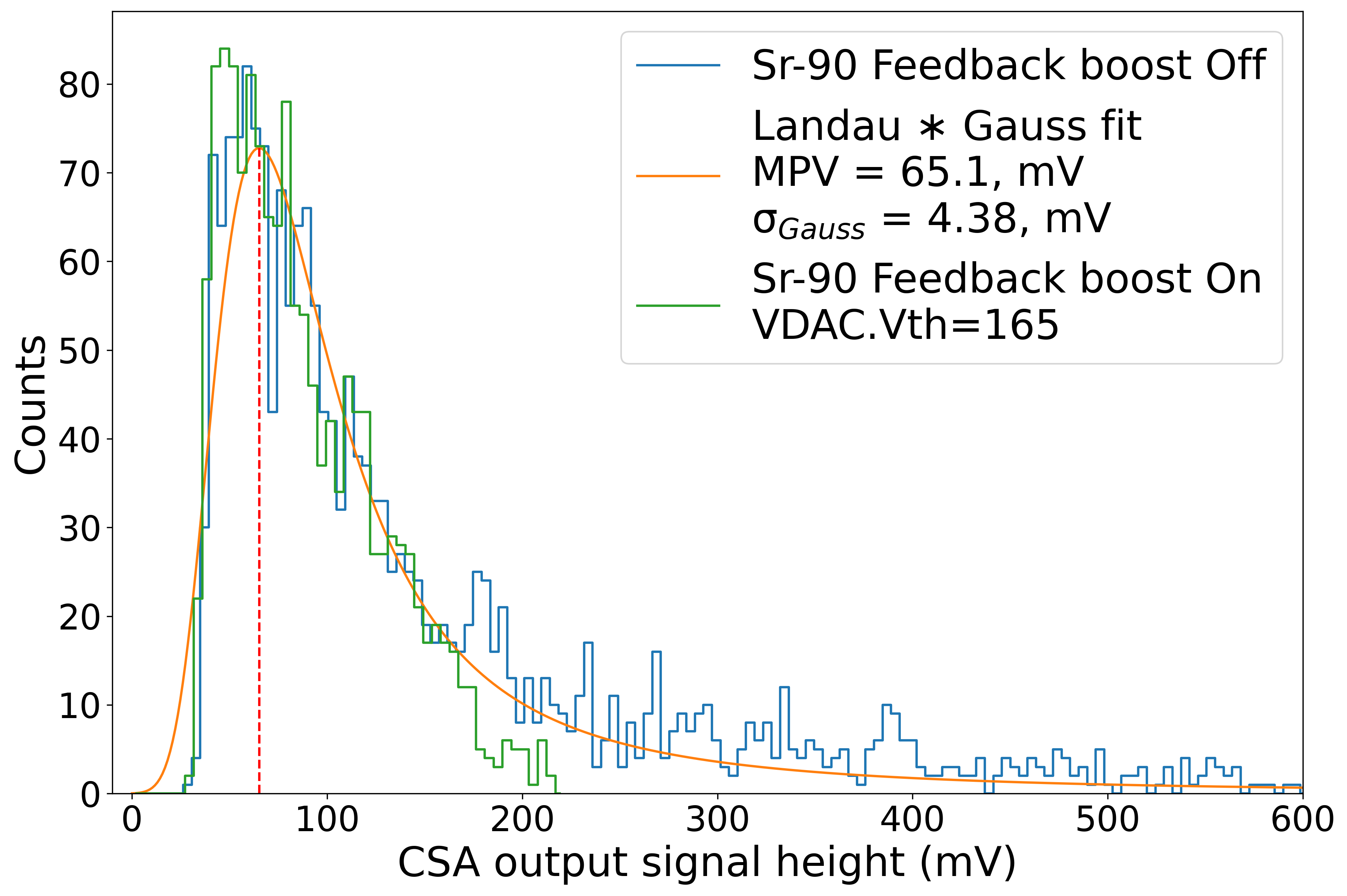}
\caption{The CSA output signal height distribution using a Sr-90 source and in case of activated feedback boost (green distribution) and when it is deactivated (blue distribution). The orange line represents a fit to the measurement data with feedback boost disabled with a convolution of Landau and Gaussian functions. The red dashed line represents the most probable value. The sensor bias voltage was set to –90~V.}
\label{fig_sig_amp_distr}
\end{figure}

The height of each output signal distribution
is shown in~Fig.~\ref{fig_sig_amp_distr}. The measurements were done with enabled (green line) and disabled (blue line) feedback boost circuit. The high energy tail of the signal height distribution is limited with enabled feedback boost circuit, because when the CSA output signal height reaches the in-pixel comparator threshold value, the feedback boost circuit starts to discharge the CSA feedback capacitor as explained in section~\ref{pixel_elec_section}, as a consequence the measured signal height becomes distorted, since a large discharge current is fed into the feedback circuit with a switch (these distorted signals were discarded from the distribution in~Fig.~\ref{fig_sig_amp_distr}). This effect does not influence the number of detected particles, since the feedback boost is enabled once the in-pixel comparator goes into a high state.

The obtained distribution without feedback boost was fitted with the convolution of Gaussian and Landau functions, and a most probable value (MPV)~of~65.1~$\pm$~4.7~mV was extracted. Assuming a depletion depth of 55~$\upmu$m, the Geant4~\cite{Geant4} simulated MPV of the deposited charge from a Sr-90 source is around 4211~$e^{-}$. It is important to note that the simulated value only accounts for carriers generated within the depletion region. The Cadence Virtuoso simulated CSA response to an input charge close to an amplifier saturation level of 31300~$e^{-}$ is 550~mV. The MPV value from Sr-90 measurement was used to calculate the calibration factor of 64.64~$\pm$~4.65~$e^{-}\cdot$mV$^{-1}$. It is in agreement with the Cadence simulation result of 56.90~$\pm$~0.94~$e^{-}\cdot$mV$^{-1}$. The experimental factor was used to calibrate the CSA output.

\begin{figure}[!hbt]
\centering
\includegraphics[width=3.2in]{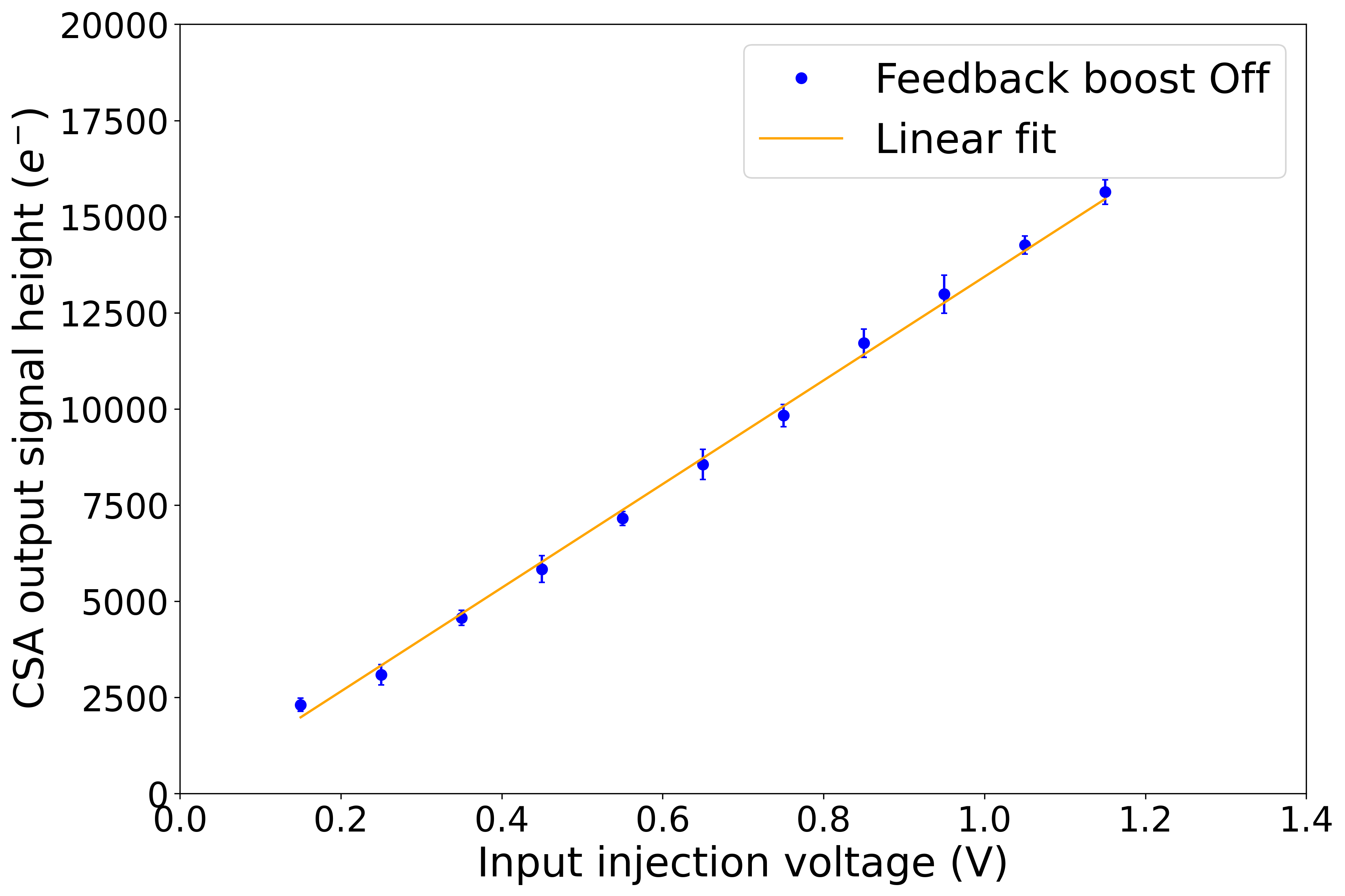}
\caption{The output characteristic of the pixel CSA, fitted with a linear function after the input injection voltage was corrected. The error bars represent the
standard deviation of measured signal heights. The sensor bias voltage was set to –90~V.}
\label{fig_csa_gain}
\end{figure}

The GECCO board provides injection circuits which generate a fast voltage pulse with controlled amplitude, duration, and number of pulses~\cite{Ehrler2021, Schimassek2021}, which are applied to the injection pad of the sensor. By varying the injection pulse amplitude and recording the CSA outputs, the output characteristic of the CSA was obtained. The measurement results are shown in~Fig.~\ref{fig_csa_gain}.

A satisfactory linearity is observed over a wide range of injection voltages. The slope of the fitted curve is 13477~$\pm$~238~$e^{-}\cdot$V$^{-1}$. The presented curve was corrected with respect to the input injection voltage offset of $-0.153$~V, which translates to a negative intercept of $-2062$~$e^{-}$. This negative intercept was consistent with observations that no measurable output signal was detected for injection voltages below approximately 0.15~V. In the following analysis of the charge injection measurements, the corrected calibration curve was used to calibrate the effective input injection voltage to the injected charge.

\subsection{Charge injection measurement}
\begin{figure}[!hbt]
\centering
\includegraphics[width=3.2in]{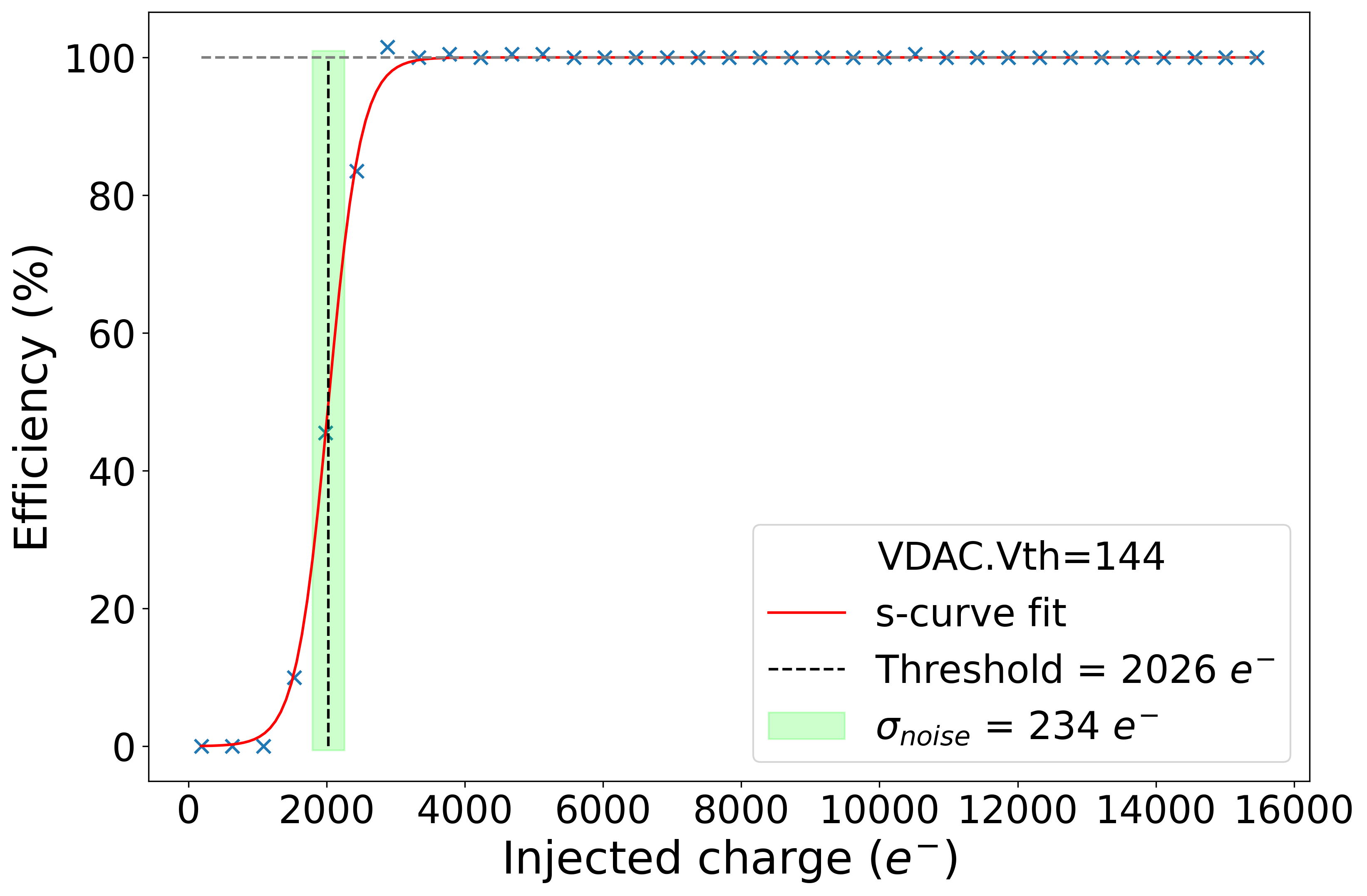}
\caption{Typical response curve of one pixel to injected pulses with different amplitudes. The distribution is fitted with an error function. The black dashed line represents the pixel threshold value, and the green bands represent the pixel noise. The sensor bias voltage was set to –90~V.}
\label{fig_s_curve}
\end{figure}

During the charge injection measurement, the response and noise of each pixel can be estimated. A fixed number of injection pulses with known amplitude was applied to each pixel and the counter value of the pixel was read out. The detection efficiency is defined as the ratio of the pixel’s counts to the number of injected pulses. During the charge injection measurement, the pixel threshold VDAC.Vth, which is common to all pixels and generated by the VDAC block, can be scanned. An example of a typical pixel response curve is shown in~Fig.~\ref{fig_s_curve}.

The distribution follows an error-function. The pixel threshold is the value of injected charge corresponding to 50\% efficiency. The width of the transition region represents the pixel noise.

\subsection{Pixel threshold tuning} \label{pix_th_tuning}
A HitPix3 pixel has a 5-bit RAM cell which can be used for the individual threshold tuning of the pixel. Four bits of the RAM cell set the tune DAC value. The output voltage range of the tune DAC block is controlled by the IPDAC value associated to the VDAC setting, which is common to the entire sensor. The IPDAC parameter can take a value in the range from~0~to~63, which corresponds to the output voltage range of the tune DAC block from~150~mV~to~300~mV. The pixel threshold tuning procedure consists of several charge injection measurements at fixed Vth and IPDAC values.

Optimal threshold values for each pixel were found with the algorithm~\ref{alg:alg_dac_tune}, depicted hereafter. The main idea of the algorithm is to determine the linear function parameters $\mathbf{slope_{i}}$ and $\mathbf{inter_{i}}$ of each pixel with number $i$, which bind the pixel threshold $\mathbf{thr_{i}}$ and RAM values and use them to reduce the pixel-to-pixel threshold dispersion. First loop, defined in lines 1-5, iterates over $q$ pixel RAM values from~$0000_2$~to~$1111_2$. In each iteration the $q$ value is uploaded to each pixel $\mathbf{RAM_{i}}$ block (line 2), charge injection measurement is performed (line 3) and each pixel $\mathbf{thr_{i}}$($q$) is extracted from the pixel response curve (line 4). Next loop, defined in lines 6-9, iterates over pixel number $i$ from~1~to~$N_{pixels}$. In each iteration $\mathbf{slope_{i}}$ and $\mathbf{inter_{i}}$ values are determined from the linear regression on sets $q$ and $\mathbf{thr_i}(q)$ (line 7), the pixel threshold value $\mathbf{thr^{'}_{i}}$ in the middle of RAM range ($q = 0111_2$) is calculated (line 8). In the next step the target threshold $\mathbf{thr_{target}}$ value is determined as a median value of the $\mathbf{thr^{'}_{i}}$ set (line 10). A typical $\mathbf{thr^{'}_{i}}$ distribution has a Gaussian shape and setting the median value as a target threshold allows to effectively use half of the RAM range ($q \leq 0111_2$) to tune pixels with $\mathbf{thr^{'}_{i}} \geq \mathbf{thr_{target}}$ and another half ($q > 0111_2$) to tune pixels with $\mathbf{thr^{'}_{i}} < \mathbf{thr_{target}}$. Next, the optimal RAM value $q^{'}_{i}$ is calculated for each pixel based on $\mathbf{thr_{target}}$, $\mathbf{slope_{i}}$ and $\mathbf{inter_{i}}$ (line 11). The optimal value is rounded to the nearest integer number. In the last step, $q^{'}_{i}$ is clipped to be within the RAM range from~$0000_2$~to~$1111_2$ and uploaded to each pixel (line 12).

\begin{algorithm}[H]
\caption{Pixel threshold tuning}\label{alg:alg_dac_tune}

  \begin{algorithmic}[1]
      \For{$q \gets 0000_2$ to $1111_2$}
      \State $ \mathbf{RAM}_\mathbf{i} \gets q \textbf{ for }  \mathbf{i} = 1,...,N_{pixels}$
      \State $ \textbf{Perform charge injection measurement} $
      \State $ \textbf{Extract pixel threshold } \mathbf{thr_i}(q) \textbf{ for }  \mathbf{i} = 1,...,N_{pixels}$
      \EndFor
      
      \For{$i \gets 1$ to $N_{pixels}$}
      \State $ \mathbf{slope}_\mathbf{i}, \mathbf{inter}_\mathbf{i} \gets \mathbf{linear fit}(q, \mathbf{thr_i}(q)) $
      \State $ \mathbf{thr^{'}_{i}} \gets \mathbf{slope}_\mathbf{i}\cdot0111_2 +  \mathbf{inter}_\mathbf{i} $
      \EndFor
      \State $ \mathbf{thr_{target}} \gets \mathbf{median}(\mathbf{thr^{'}_{i}}) $
      \State $\mathbf{q^{'}_{i}} \gets \mathbf{ceil}( \frac{\mathbf{thr_{target}} - \mathbf{inter_{i}}}{\mathbf{slope_{i}}}) \textbf{ for }  \mathbf{i} = 1,...,N_{pixels}$
      
      \State $ \mathbf{RAM}_\mathbf{i} \gets \mathbf{max}(0000_{2}, \mathbf{min}(\mathbf{q^{'}_{i}}, 1111_{2})) \textbf{ for }  \mathbf{i} = 1,...,N_{pixels}$
  \end{algorithmic}
\end{algorithm}

The duration of the pixel threshold tuning procedure depends on two factors: the number of $q$ values used for the fit of pixel linear functions and the total number of pixels. In the algorithm~\ref{alg:alg_dac_tune} the most time consuming part is the loop over $q$ values (one step takes approximately 20 seconds). Typically, the pixel threshold tuning procedure requires five $q$ values to reliably determine the values of $\mathbf{slope_{i}}$ and $\mathbf{inter_{i}}$ of a pixel. For the clinical application, the pixel threshold tuning procedure is foreseen to be performed once a day.

\begin{figure}[!hbt]
\centering
\includegraphics[width=3.2in]{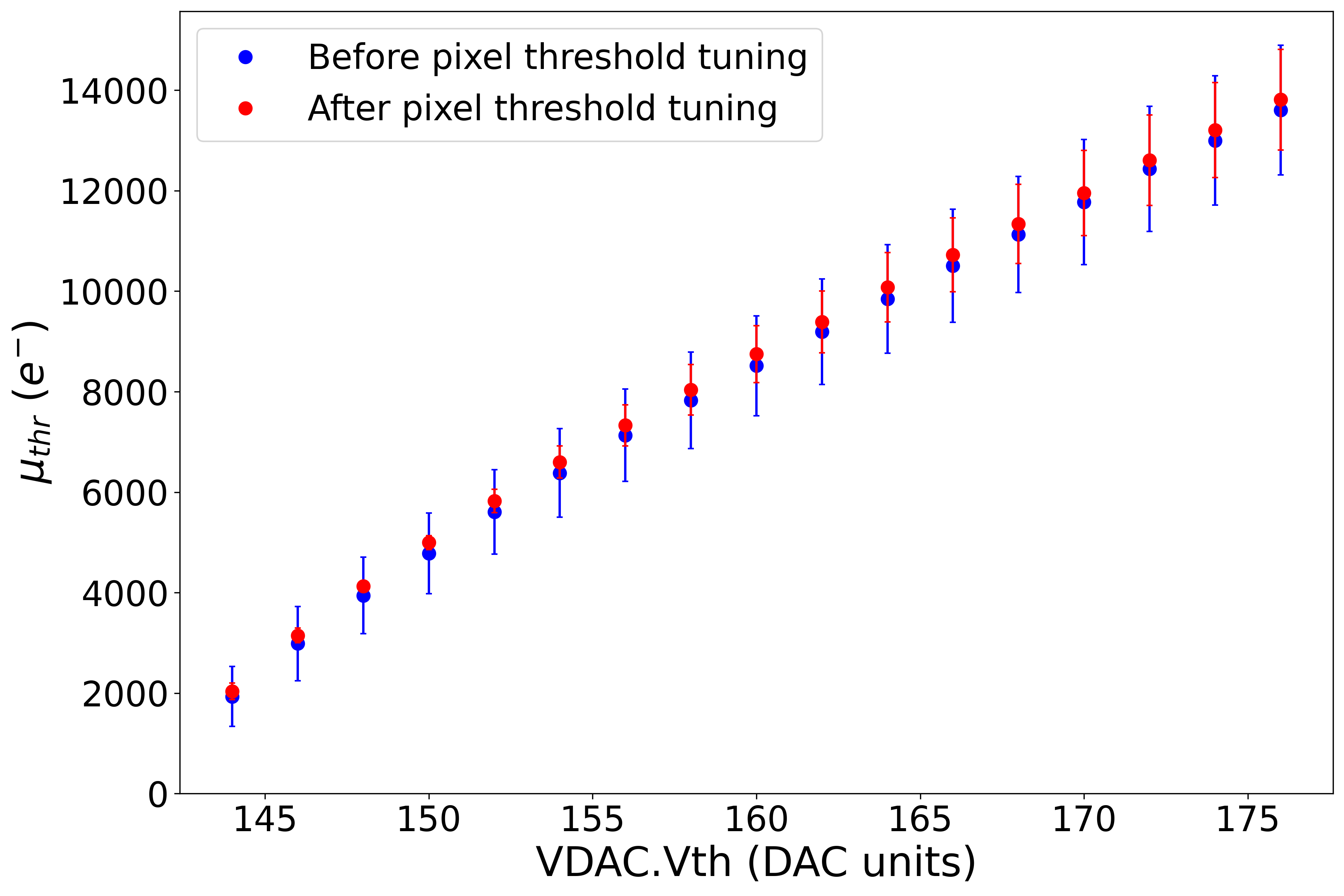}
\caption{Mean ($\upmu_{thr}$) of the pixel threshold distribution before (blue points) and after (red points) pixel threshold tuning extracted from a Gaussian fit at different VDAC.Vth values. The VDAC.IPDAC was set to 15. The error bars stand for the standard deviations extracted from a Gaussian fit. The sensor bias voltage was set to –90~V.}
\label{fig_th_mean_std_first_case}
\end{figure}

\begin{figure}[!hbt]
\centering
\includegraphics[width=3.2in]{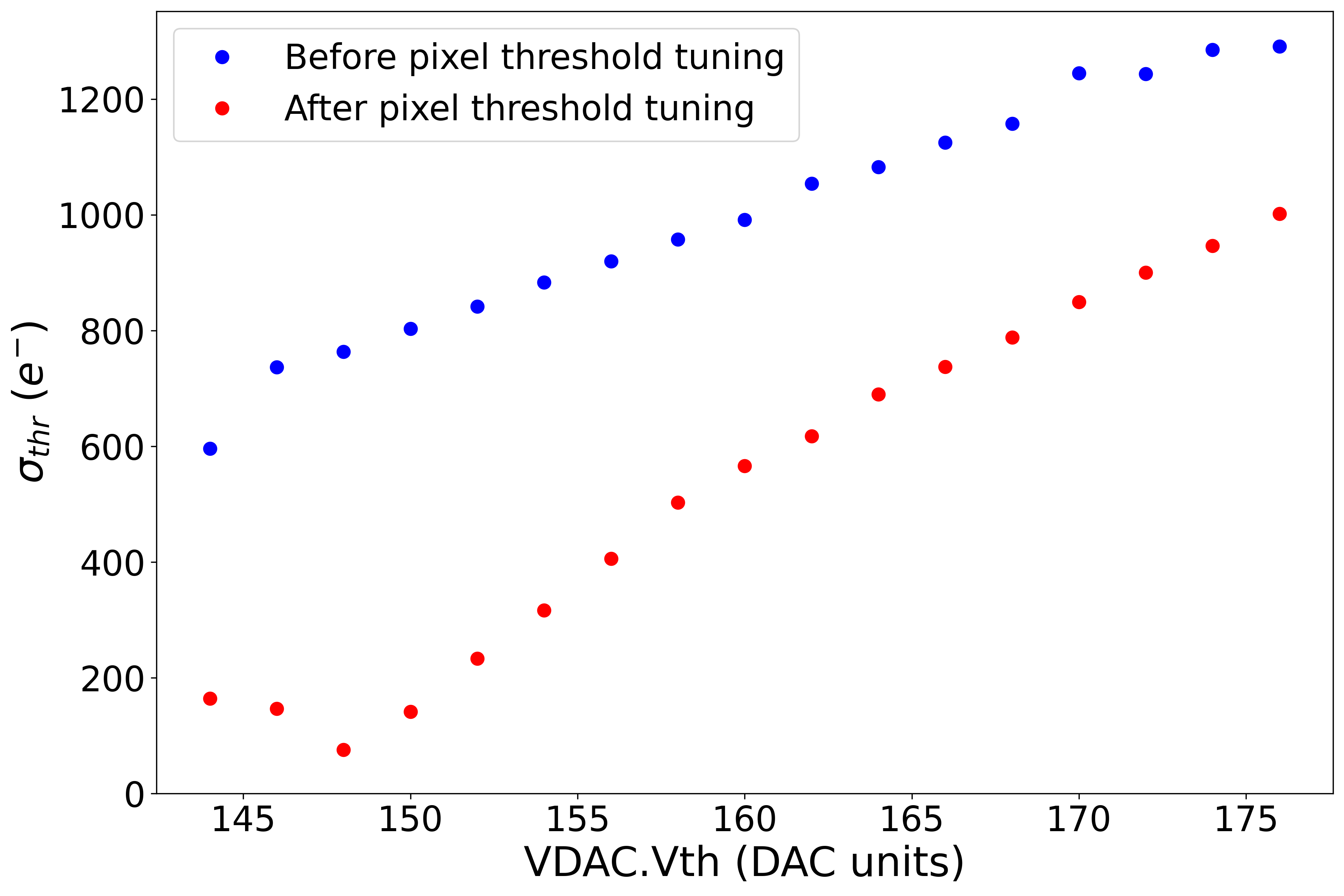}
\caption{Standard deviation ($\sigma_{thr}$) of the pixel threshold distribution before (blue points) and after (red points) pixel threshold tuning extracted from a Gaussian fit at different VDAC.Vth values. The VDAC.IPDAC was set to 15. The sensor bias voltage was set to –90~V.}
\label{fig_th_mean_std_second_case}
\end{figure}

Charge injection measurements were performed at different VDAC.Vth values before and after pixel threshold tuning. All injection voltages were converted to the equivalent charge by using the charge calibration data presented above. The distributions of pixel thresholds were fitted with a Gaussian function. The mean and standard deviation of these thresholds, extracted from a Gaussian fit at various VDAC.Vth values, are shown in Fig.~\ref{fig_th_mean_std_first_case} and Fig.~\ref{fig_th_mean_std_second_case}, respectively.

Fig.~\ref{fig_th_mean_std_first_case} shows that the dependence of the mean of the pixel threshold distribution on VDAC.Vth is not affected by the pixel threshold tuning procedure, as expected. As shown by Fig.~\ref{fig_th_mean_std_second_case}, the standard deviation of the pixel threshold distribution is reduced by more than a factor of four after the pixel threshold tuning in a range of VDAC.Vth between 146 and 150 DAC units. At HIT treatment conditions, the proton beam with energy 221.06~MeV provides the minimal energy deposition of all beam settings.
Geant4 simulations of the energy deposition in HitPix3 at -90~V bias voltage were performed with a 221.06~MeV proton beam. 
The resulting Landau distribution of the generated signals has a low-signal tail below the MPV from 5700~$e^{-}$ to 8800~$e^{-}$. Therefore, a VDAC.Vth threshold of 148~DAC~units (corresponding to $\upmu_{thr}$~=~4132~$e^{-}$ and $\sigma_{thr}$~=~76~$e^{-}$) is sufficient to cover the whole range of the expected signals at HIT conditions, since all other beam settings provide deposited signals above 5700~$e^{-}$.
 
The mean ENC ($\upmu_{ENC}$) of HitPix3 is equal to 162~$\pm$~22~$e^{-}$. As anticipated, $\upmu_{ENC}$ was found to be insensitive to VDAC.Vth when varying the latter from 144 to 176~DAC~units, since it depends only on the electrical pixel properties.

\subsection{Counter and adder readout}
\begin{figure}[!hbt]
\centering
\includegraphics[width=3.4in]{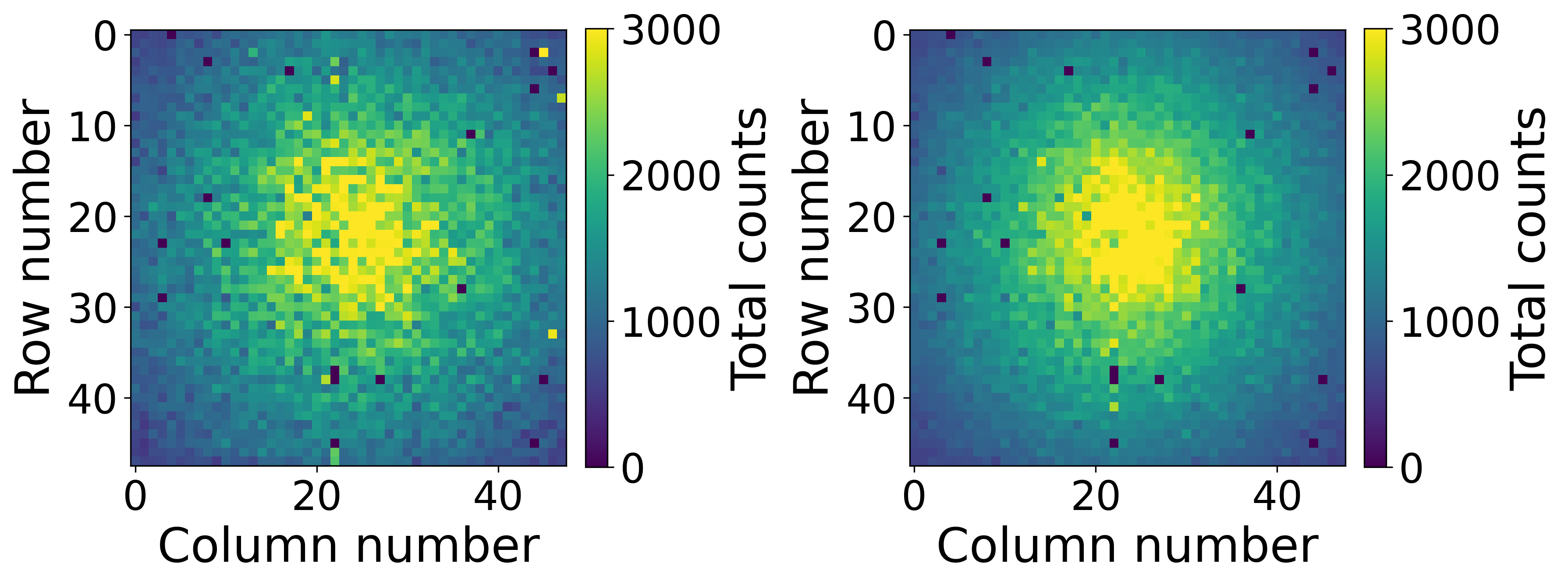}
\caption{Cumulative spot profile observed when illuminating the sensor with an Sr-90 source, in counter mode before (left) and after (right) the pixel threshold tuning. The sensor bias voltage was set to –90~V. Frame duration: 5~ms. Number of frames: 3000. VDAC.Vth~=~144~DAC~units. Readout speed: 180~Mbit/s.}
\label{fig_counters}
\end{figure}

The counter and adder readout modes of the HitPix3 sensor were tested with a Sr-90 source. The bias voltage of the sensor was set to –90~V, the frame time window to 5~ms and the readout speed to 180~Mbit/s. The frame time window can take values from~1~$\upmu$s~to~50~ms with 1~$\upmu$s steps. 3000~frames were recorded for each mode before and after pixel threshold tuning. The pixels were marked as noisy and masked if they had a hit rate greater than 0.01 counts per frame at given comparator threshold value in a measurement without a Sr-90 source. Eighteen pixels were found noisy, i.e. 0.7\% of all pixels in the sensor. The comparator threshold value VDAC.Vth was set to 144~DAC~units (corresponding to $\upmu_{thr}$~=~2038~$e^{-}$ and $\sigma_{thr}$~=~164~$e^{-}$), which is lower than the MPV estimated with Sr-90 source. Thresholds below 144~DAC~units increase the fraction of noisy pixels to 100\%. The used threshold value cuts part of the low signal range of the Landau distribution generated by the Sr-90 source, but it is sufficient for the detection of the minimal generated signal in HIT conditions (a detailed explanation is provided in subsection~\ref{pix_th_tuning}). Fig.~\ref{fig_counters} shows the cumulative spot profile of the Sr-90 source measured in counter mode before (left plot) and after (right plot) the pixel threshold tuning.

\begin{figure}[!hbt]
\centering
\includegraphics[width=3.3in]{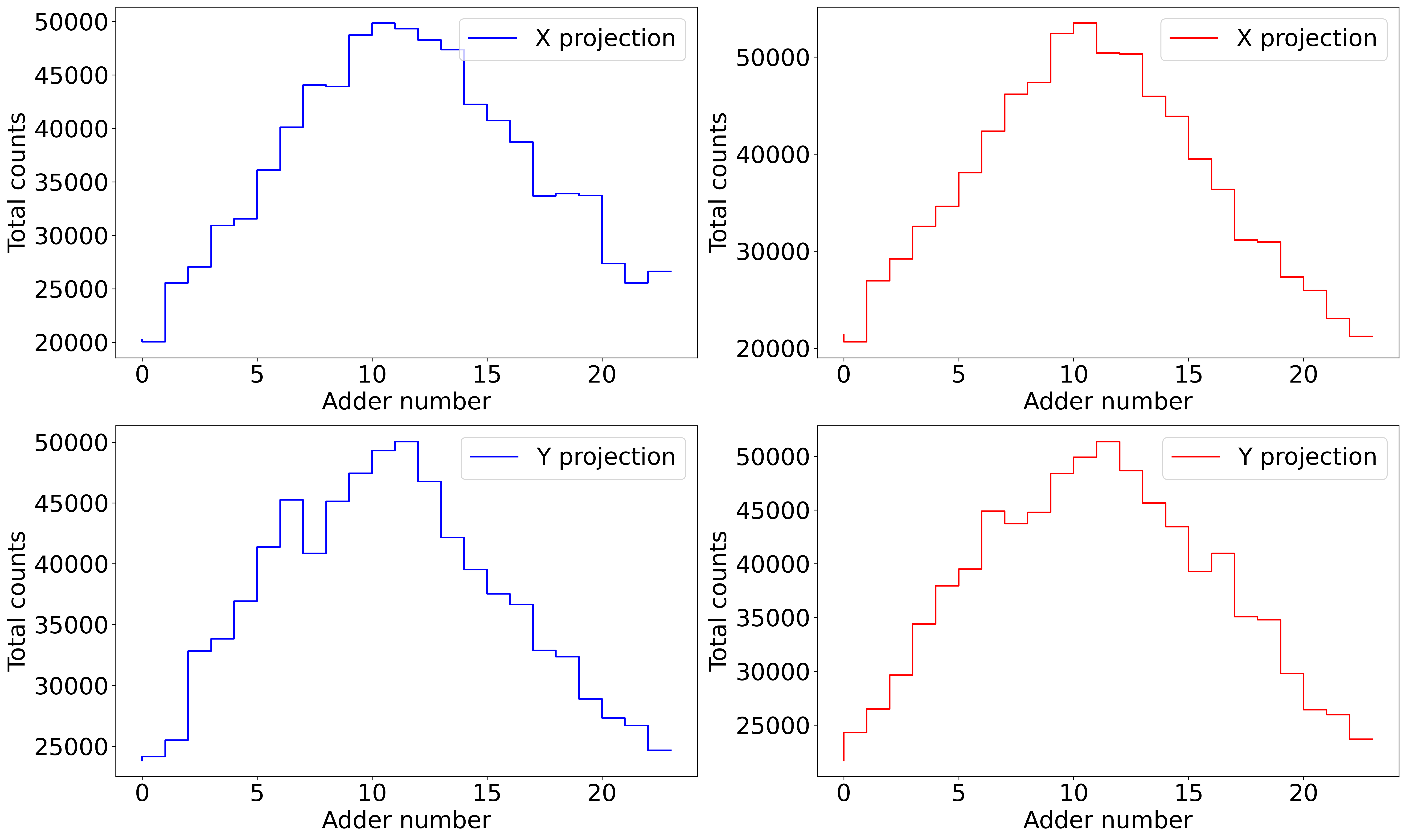}
\caption{Cumulative $x$- (top) and $y$- (bottom) projections of the Sr-90 spot profile measured in adder mode before (blue lines) and after (red lines) the pixel threshold tuning. The sensor bias voltage was set to –90~V. Frame duration: 5~ms. Number of frames: 3000. VDAC.Vth~=~144~DAC~units. Readout speed: 180~Mbit/s.}
\label{fig_adders}
\end{figure}

One observes that the measured spot profile gains in regularity and contour smoothness after the pixel threshold tuning  as a consequence of the substantial reduction of the pixel-to-pixel threshold dispersion, therefore most of the pixels react to the incoming particles in the same way. The same effect can be seen in the cumulative $x$- and $y$- projections of the spot profile measured in adder mode as is shown in~Fig.~\ref{fig_adders}. In counter mode, the dead time required to shift out one row of data is equal to 7~$\upmu$s at 180~Mbit/s readout speed. In adder mode, the pipeline readout can be used, i.e. the previous frame is read out while the current frame is being recorded. In this case the dead time will be $\sim$2~$\upmu$s, caused by the control signals which are used for the data transfer from the adders to the shift register.

\subsection{Pixel masking}
\begin{figure}[!hbt]
\centering
\includegraphics[width=2.2in]{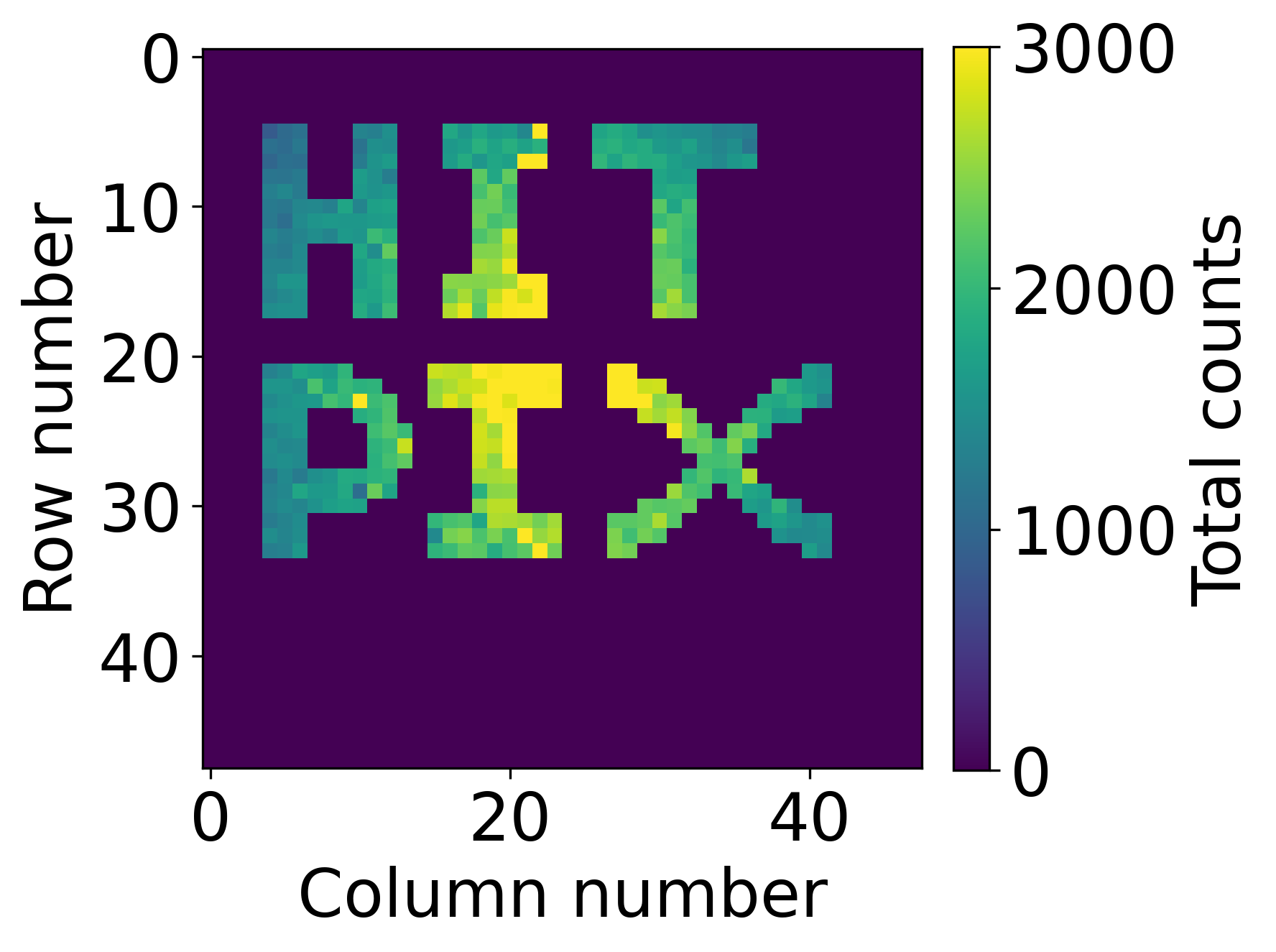}
\caption{Cumulative Sr-90 spot profile measured in counter mode with a masking pattern applied. The sensor bias voltage was set to –90~V. Frame duration: 5~ms. Number of frames: 3000. VDAC.Vth~=~144~DAC~units. Readout speed: 180~Mbit/s.}
\label{fig_hitpix_logo}
\end{figure}

The last bit Q$\langle$4$\rangle$ of the 5-bit pixel RAM is used for the purpose of masking the pixel output. To check the functionality of this bit several mask patterns were generated and uploaded to the sensor. An example of a cumulative Sr-90 profile measured in counter mode with masked pixels is shown in~Fig.~\ref{fig_hitpix_logo}.

The measurement confirms the masking functionality. It will be used to mask damaged pixels or pixels with very high noise. This function becomes especially useful in adder mode to alleviate the impact of high radiation doses on the sensor performance.

\section{Conclusion}
This article describes the new sensor system HitPix3 based on hit counting HV-CMOS sensors, which targets the use as a beam monitor for ion beam therapy. The sensor is tailored to the requirements of the beam monitoring system at the Heidelberg Ion Beam Therapy center.

The monolithic active pixel sensor HitPix3 is produced with commercial HV-CMOS technology on high-resistivity substrate with counting pixel electronics and frame-based readout. The pixel size is adapted for the beam position and size resolution requirements and the sensor area of multi-sensor readout modules can fulfill the large sensitive area requirement of a beam monitor. The sensor features an integrated hit counting capability as well as a built-in column- and row-wise projection functionality, allowing it to process high particle rates with fast readout at minimum latency.

Several new designs were presented in this article, including a modified amplifier, a second projection adder, in-pixel threshold tuning capacity and a faster readout concept.

The presented laboratory measurements of the unirradiated sensors confirm the full functionality of all the design features introduced in HitPix3. The tested sensors could withstand a high bias voltage up to –180~V without breakdown and were read out at the maximum readout speed of the current firmware of 190~Mbit/s. Furthermore, the new in-pixel threshold tuning feature allowed to reduce the pixel-to-pixel variation in the threshold distribution by more than a factor of four. This resulted in a more uniform response across all pixels, improving the regularity and contour smoothness of measured Sr-90 source spot profiles.

The next planned tests will focus on the detection performance of high intensity ion beams and sensor radiation tolerance.
The tests will also address the readout of multi-sensor assemblies as required for the monitoring of the larger beam spots at HIT (FWHM of 20~mm for carbon ions and 30~mm for protons). The outcome of these tests will allow to optimize the design of the next sensor, which will be a full size, 2~cm$~\times~$2~cm large die to cover the required 26~cm$~\times~$26~cm sensitive area with a 13$~\times~$13 sensor matrix. The next sensor is supposed to incorporate all features needed for its application as a beam monitor at HIT.

\section*{CRediT authorship contribution statement}
\textbf{Hui Zhang:} Conceptualization, Methodology, Investigation, Visualization, Writing~-~original draft. \textbf{Bogdan Topko:} Data curation, Formal analysis, Methodology, Investigation, Software, Validation, Visualization, Writing~-~original draft. \textbf{Ivan Peri\'c:} Conceptualization, Supervision, Resources, Funding acquisition.

\section*{Acknowledgments}
The authors would like to thank their colleagues F.~Ehrler for the help in the sensor configuration and the PCB design verification, U.~Husemann and A.~Dierlamm for the project administration, funding acquisition and the article text reviewing, B. Regnery for the article text reviewing, H.~J.~Simonis and W.~Rehm for the sensor wire bonding.

This work was supported by HEiKA - Heidelberg Karlsruhe Strategic Partnership, Heidelberg University, Karlsruhe Institute of Technology (KIT); Germany. This work was also supported by Heidelberg Ion Beam Therapy Center (HIT). The authors acknowledge financial support through the German Federal Ministry of Education and Research (BMBF) within the ARTEMIS project (Grant number: 13GW0436A).

\section*{Declaration of competing interest}
The authors declare that they have no known competing financial interests or personal relationships that could have appeared to influence the work reported in this paper.

\section*{Data availability}
Data will be made available on request.

 \bibliographystyle{elsarticle-num}

\end{document}